\documentclass[runningheads]{llncs}
\usepackage{cite}
\usepackage{amsmath,amssymb,amsfonts}
\usepackage{algorithmic}
\usepackage{graphicx}
\usepackage{textcomp}
\usepackage{xcolor}
\usepackage{bmpsize}
\usepackage{float}
\usepackage{lipsum}
\usepackage{appendix}

\usepackage[colorlinks=true,urlcolor=black]{hyperref}

\makeatletter
\newcommand\notsotiny{\@setfontsize\notsotiny\@vipt\@viipt}
\makeatother
\begin{document}
\title{Can process mining help in anomaly-based intrusion detection?}
%
%
\author{Yinzheng Zhong\orcidID{0000-0001-8477-3956} \and
Alexei Lisitsa\orcidID{0000-0002-3820-643X}}
%
%
\institute{Department of Computer Science, University of Liverpool, UK \email{y.zhong10@liverpool.ac.uk,
a.lisitsa@liverpool.ac.uk}\\
}
\maketitle              
\begin{abstract}
    In this paper, we consider the naive applications of process mining in network traffic comprehension, traffic anomaly detection, and intrusion detection. We standardise the procedure of transforming packet data into an event log. We mine multiple process models and analyse the process models mined with the inductive miner using ProM \cite{verbeek2010prom} and the fuzzy miner using Disco \cite{Gnther2012DiscoDY}. We compare the two types of process models extracted from event logs of differing sizes. We contrast the process models with the RFC TCP state transition diagram \cite{postel1981transmission} and the diagram \cite{bishop2005tcp} by Bishop et al. We analyse the issues and challenges associated with process mining in intrusion detection and explain why naive process mining with network data is ineffective.

\keywords{
Anomaly detection \and Intrusion detection \and Process mining \and Cyber security.}
\end{abstract}
\section{Introduction}
    This paper contains some of our earlier works prior to \cite{zhong2022process}. Our objective is to create an intrusion detection system using process mining (PM) and process models. The models can then be utilised for analytics \cite{van2004process, van2011process}. Network intrusion detection systems (NIDS) are typically based on the following methodologies. First, the features are extracted from the flow-level or connection-level data, and then the features are generated from statistical data, such as the number of packets and the average payload size \cite{lee1998data, hsu2019toward}. For classification, these approaches frequently employ data mining, machine learning, and other statistical methods. Second, audit trail methods examine the network's activities, such as the payload content, flag set, or numerical values \cite{roesch1999snort, wespi2000intrusion}. The issue with the majority of existing techniques is that they are unable to observe the global process structure of a communication.
    
    The paper by Van der Aalst and de Medeiros \cite{van2005process} proposed to use process mining on audit trails and conformance checking for anomaly detection; however, additional research for NIDS is required. Our experiment is predicated on the premise that the process models can be mined from event logs using PM techniques, and then new cases will be validated using conformance checking. The anomaly will then be identified based on the conformance score. We complied with the directive by converting network packet data to event logs, which may facilitate PM adaptation.
    
    We employ ProM for inductive mining algorithm, which is related to \cite{van2005process}. The inductive miner will mine models in the form of Petri nets. The extracted models will then be utilised for additional analysis, such as conformance checking. In contrast, we will compare fuzzy mining as an alternative PM algorithm \cite{gunther2007fuzzy}.
    
    The paper is structured as follows. First, we will look at related works that use similar approaches; second, we will demonstrate the dataset we use and how the packet data are converted into the event log; then, we will discuss the experimental setups and compare the models. After that, we show the result of conformance checking and discuss the reason for the inefficiency of PM for IDS. Finally, we introduce a novel design of a preprocessing technique that can be used in NIDS which is inspired by PM.
    
\section{Related Works}
    The technique introduced in \cite{myers2018anomaly} is used on industrial control systems (ICS) that identifies anomalous behaviours and cyber-attacks using ICS data logs and the conformance checking analysis technique. The paper demonstrated the preprocessing of ICS data logs for PM and the result of anomaly detection. The inductive mining algorithm discovers their process model, and as a result, the average F-score for detecting attacks across two datasets is 0.81. A case is categorised as an anomaly if it does not produce a fitness score of 1. The paper \cite{myers2018anomaly} is not focused on NIDS.
    
    On top of \cite{van2005process}, papers \cite{mishra2017process, mishra2018analysis} provide expanded discussions on PM for IDS, discussing potential benefits of using PM for IDS, comparing different mining algorithms (such as the heuristic miner and the $\alpha$-algorithm), and displaying cyber-attacks detected in various industrial sectors, etc. Nonetheless, the precise method for implementing PM for IDS remains undetermined.
    
    The paper \cite{9640192} uses Inductive Miner Infrequent algorithm to detect anomalous behaviors on the e-commerce platform. The data is processed based on the log recorded from the ModSecurity, which is a open source web application firewall, for the purpose of adapting PM. The events are mainly consist of URLs and activated ModSecurity rules. The provided average fitness scores show that anomalies have lower scores, however, they do not have a clear measurement for F-score or accuracy.
    
    The methodology described in \cite{ebrahim2022anomaly} is based on PM and employs social network analysis metrics to identify anomalous behaviour. Instead of performing conformance testing, this method analyses social network metrics to identify anomalies. The log of normal traces is extracted from the process model using the workflow Petri net designer (WoPeD) process modelling tool \cite{eckleder2008woped}; the role-trace access control matrix, which is based on the normal traces and the role-activity access control matrix, is then created; and finally, the social network is constructed using the role-trace access control matrix and all logs. The experiment's F-score is greater than 0.92, per the outcome.
    
    In general, PM applications in intrusion detection are not well established.
    
\section{Data Prepossessing}
    \begin{table}[H]
    \caption{The event log segment. The three zeros in flags are the reserved bits.}
    \centering
    \resizebox{\textwidth}{!}{%
    \begin{tabular}{|c|c|c|c|c|c|c|c|}
    \hline
    Case\_ID & Timestamp & Src\_IP & Dst\_IP & Scr\_port & Dst\_port & Flags & Side \\ \hline
    151043 & 2017/07/04 14:00:35.190179000 & 192.168.10.5 & 68.67.178.110 & 52892 & 80 & 000.SYN. & C \\ \hline
    151043 & 2017/07/04 14:00:35.222535000 & 68.67.178.110 & 192.168.10.5 & 80 & 52892 & 000.ACK.SYN. & S \\ \hline
    151043 & 2017/07/04 14:00:35.222586000 & 192.168.10.5 & 68.67.178.110 & 52892 & 80 & 000.ACK. & C \\ \hline
    151043 & 2017/07/04 14:00:35.237412000 & 192.168.10.5 & 68.67.178.110 & 52892 & 80 & 000.ACK.PSH. & C \\ \hline
    151043 & 2017/07/04 14:00:35.270301000 & 68.67.178.110 & 192.168.10.5 & 80 & 52892 & 000.ACK. & S \\ \hline
    151043 & 2017/07/04 14:00:35.270305000 & 68.67.178.110 & 192.168.10.5 & 80 & 52892 & 000.ACK. & S \\ \hline
    151043 & 2017/07/04 14:00:35.467062000 & 68.67.178.110 & 192.168.10.5 & 80 & 52892 & 000.ACK. & S \\ \hline
    151043 & 2017/07/04 14:00:35.467285000 & 68.67.178.110 & 192.168.10.5 & 80 & 52892 & 000.ACK.PSH. & S \\ \hline
    151043 & 2017/07/04 14:00:35.467347000 & 192.168.10.5 & 68.67.178.110 & 52892 & 80 & 000.ACK. & C \\ \hline
    151043 & 2017/07/04 14:00:35.467352000 & 68.67.178.110 & 192.168.10.5 & 80 & 52892 & 000.ACK.PSH. & S \\ \hline
    151043 & 2017/07/04 14:00:35.511515000 & 192.168.10.5 & 68.67.178.110 & 52892 & 80 & 000.ACK. & C \\ \hline
    151043 & 2017/07/04 14:00:45.461285000 & 192.168.10.5 & 68.67.178.110 & 52892 & 80 & 000.ACK. & C \\ \hline
    151043 & 2017/07/04 14:00:45.466144000 & 68.67.178.110 & 192.168.10.5 & 80 & 52892 & 000.ACK.FIN. & S \\ \hline
    151043 & 2017/07/04 14:00:45.466173000 & 192.168.10.5 & 68.67.178.110 & 52892 & 80 & 000.ACK. & C \\ \hline
    151043 & 2017/07/04 14:00:45.466198000 & 192.168.10.5 & 68.67.178.110 & 52892 & 80 & 000.ACK.FIN. & C \\ \hline
    151043 & 2017/07/04 14:00:45.493763000 & 68.67.178.110 & 192.168.10.5 & 80 & 52892 & 000.ACK. & S \\ \hline
    151043 & 2017/07/04 14:00:45.498542000 & 68.67.178.110 & 192.168.10.5 & 80 & 52892 & 000.ACK. & S \\ \hline
    281008 & 2017/07/05 17:03:45.498235000 & 192.168.10.16 & 23.194.141.47 & 51226 & 443 & 000.SYN. & C \\ \hline
    281008 & 2017/07/05 17:03:45.521284000 & 23.194.141.47 & 192.168.10.16 & 443 & 51226 & 000.ACK.SYN. & S \\ \hline
    281008 & 2017/07/05 17:03:45.521360000 & 192.168.10.16 & 23.194.141.47 & 51226 & 443 & 000.ACK. & C \\ \hline
    281008 & 2017/07/05 17:03:45.521561000 & 192.168.10.16 & 23.194.141.47 & 51226 & 443 & 000.ACK.PSH. & C \\ \hline
    281008 & 2017/07/05 17:03:45.544708000 & 23.194.141.47 & 192.168.10.16 & 443 & 51226 & 000.ACK. & S \\ \hline
    281008 & 2017/07/05 17:03:45.545025000 & 23.194.141.47 & 192.168.10.16 & 443 & 51226 & 000.ACK.PSH. & S \\ \hline
    281008 & 2017/07/05 17:03:45.545073000 & 192.168.10.16 & 23.194.141.47 & 51226 & 443 & 000.ACK. & C \\ \hline
    281008 & 2017/07/05 17:03:45.561805000 & 192.168.10.16 & 23.194.141.47 & 51226 & 443 & 000.ACK.PSH. & C \\ \hline
    281008 & 2017/07/05 17:03:45.624241000 & 23.194.141.47 & 192.168.10.16 & 443 & 51226 & 000.ACK. & S \\ \hline
    281008 & 2017/07/05 17:03:45.820846000 & 192.168.10.16 & 23.194.141.47 & 51226 & 443 & 000.ACK.PSH. & C \\ \hline
    281008 & 2017/07/05 17:03:45.843822000 & 23.194.141.47 & 192.168.10.16 & 443 & 51226 & 000.ACK. & S \\ \hline
    281008 & 2017/07/05 17:03:45.921777000 & 23.194.141.47 & 192.168.10.16 & 443 & 51226 & 000.ACK.PSH. & S \\ \hline
    281008 & 2017/07/05 17:03:45.964684000 & 192.168.10.16 & 23.194.141.47 & 51226 & 443 & 000.ACK. & C \\ \hline
    281008 & 2017/07/05 17:03:46.802581000 & 192.168.10.16 & 23.194.141.47 & 51226 & 443 & 000.ACK.PSH. & C \\ \hline
    281008 & 2017/07/05 17:03:46.825827000 & 23.194.141.47 & 192.168.10.16 & 443 & 51226 & 000.ACK. & S \\ \hline
    281008 & 2017/07/05 17:03:46.827025000 & 23.194.141.47 & 192.168.10.16 & 443 & 51226 & 000.ACK.PSH. & S \\ \hline
    281008 & 2017/07/05 17:03:46.827041000 & 192.168.10.16 & 23.194.141.47 & 51226 & 443 & 000.ACK. & C \\ \hline
    281008 & 2017/07/05 17:03:53.457141000 & 192.168.10.16 & 23.194.141.47 & 51226 & 443 & 000.ACK.FIN. & C \\ \hline
    281008 & 2017/07/05 17:03:53.480155000 & 23.194.141.47 & 192.168.10.16 & 443 & 51226 & 000.ACK.PSH. & S \\ \hline
    281008 & 2017/07/05 17:03:53.480156000 & 23.194.141.47 & 192.168.10.16 & 443 & 51226 & 000.ACK.FIN. & S \\ \hline
    281008 & 2017/07/05 17:03:53.480264000 & 192.168.10.16 & 23.194.141.47 & 51226 & 443 & 000.RST. & C \\ \hline
    281008 & 2017/07/05 17:03:53.480268000 & 192.168.10.16 & 23.194.141.47 & 51226 & 443 & 000.RST. & C \\ \hline
    \end{tabular}
    }
    \label{tab:event_log_example}
    \end{table}
    
    In our experiment, we utilised the IDS2017 dataset \cite{unb}. The dataset contains common attacks including Bruteforce, DoS, and Botnet, among others. The dataset includes PCAP packet data as well as CSV data sheets with analytical features that have been preprocessed. Focusing on packet-level detection, we will only utilise binary packet data and extract necessary data from this dataset using TShark. The processed and transformed data must then be applied to PM as event logs. In this paper, we only consider TCP data. We will keep two attributes for the event logs, the first attribute is the packet flags information; the second will be the are labels which indicate whether a packet is sent from the server (S) or the client (C).
    
    First, we filtered out invalid packets that do not use the TCP protocol or were missing data in part. Second, we reconstructed each case based on the socket pair (source IP\_Port and destination IP\_Port) and assigned a unique case ID to each flow. A \emph{case} in PM is essentially an instance of a series of actions/activities. In network traffic, we define it as a series of packets within a connection; therefore, in subsequent contexts, we may refer to the case as the network flow. The TCP header flags will then be converted into human-readable strings, such as ACK (for ACK flag sets) or ACK.FIN (for ACK and FIN flags set). We define a complete flow as one in which the SYN flag is set at the beginning and the RST or FIN flag is set at the end; thus, all cases that do not begin with SYN or do not contain RST/FIN will be discarded.
    
    The IDS2017 dataset includes numerous PCAP binary files, each of which stores packets from a specific host. Some files only contain normal traffic data, so we combine the processed event logs from these files to create a comprehensive normal event log (the complete-log). Flows involved in attacks are also transformed separately into anomalous event logs. All normal and anomalous event logs are saved as CSV data sheets. These files will constitute our final datasets, which can be fed directly into the PM tools.
    
    Table \ref{tab:event_log_example} shows an example of a segment of the event log. Note that IPs and Ports are not used in PM as they are used for determine the case ID only. The data columns used in the experiment are case IDs, timestamps, header flags and side labels. We select header flags and side labels as our attributes as they are non-numerical and the transitional information can be constructed with these data, which is suitable for PM. The transition here is defined as $(a, b)$ where $a$ and $b$ are consecutive events within a case.
    
\section{Experiment Setups}
    We retained every hyperparameter as default in both ProM and Disco in PM. The first reason is that understanding the problems is more important than looking through different mining hyperparameters. Second, we want to mine authentic models from the event logs we provide because rare cases are not necessarily anomalies; also we know these rare cases are normal as they come from the normal event log.
    
    For the purpose of comparison and comprehension, we sub-sampled the complete-log into different smaller-sized datasets and mined the process model from these subsets with both process discovery algorithms. Here are the setups for inductive miner in ProM.
    
    \begin{enumerate}
        \item Process model with 5 flows \#1.
        \item Process model with 5 flows \#2.
        \item Process model with 100 flows \#1.
        \item Process model with 100 flows \#2.
        \item Process model with 20k flows \#1. \label{model_for_comformance_checking}
        \item Process model with 20k flows \#2.
        \item Process model with 100k flows \#1.
        \item Process model with 100k flows \#2.
        \item Process model of complete set.
    \end{enumerate}
    
    Similar setups are used for fuzzy miner with Disco; however, Disco is limited to 50k cases. Subsets with fewer than 50k cases are identical to inductive miner subsets. The details are presented below.
    
    \begin{enumerate}
        \item Process model with 5 flows \#1.
        \item Process model with 5 flows \#2.
        \item Process model with 100 flows \#1.
        \item Process model with 100 flows \#2.
        \item Process model with 20k flows \#1.
        \item Process model with 20k flows \#2.
        \item Process model with 50k flows \#1.
        \item Process model with 50k flows \#2.
    \end{enumerate}
    
\section{Model Comparisons}
    \subsection{Process Models}
\begin{figure}[h]
        	\begin{center}
        	\includegraphics[width=4in]{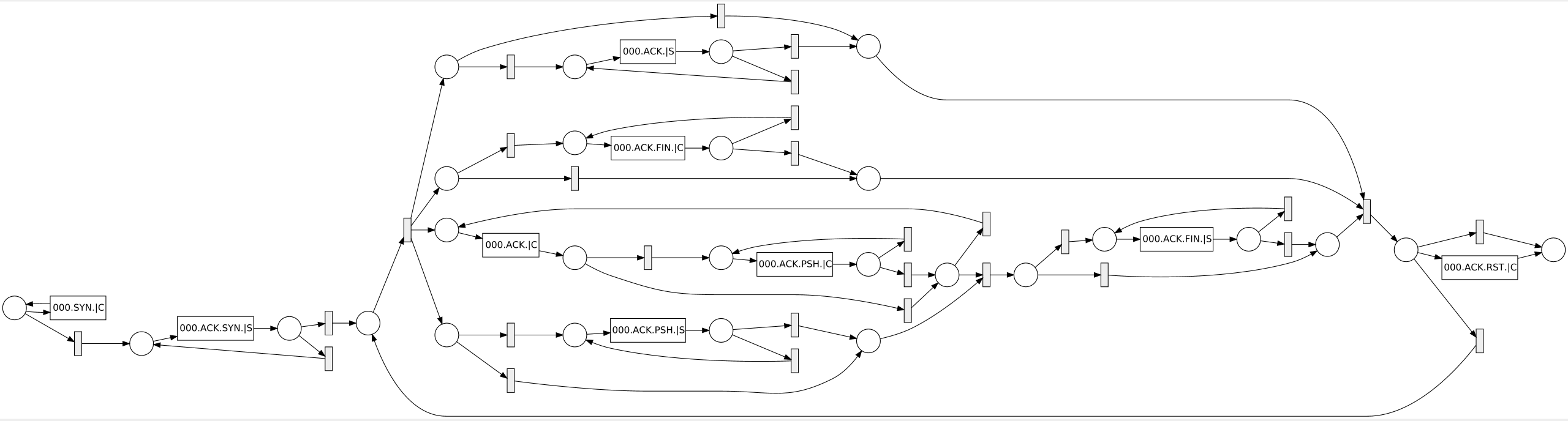}
        	\caption{The Petri net mined with 5 cases.}
        	\label{fig:inductive_model_compare_5}
        	\end{center}
        \end{figure}

\begin{figure}[h]
        	\begin{center}
        	\includegraphics[width=4in]{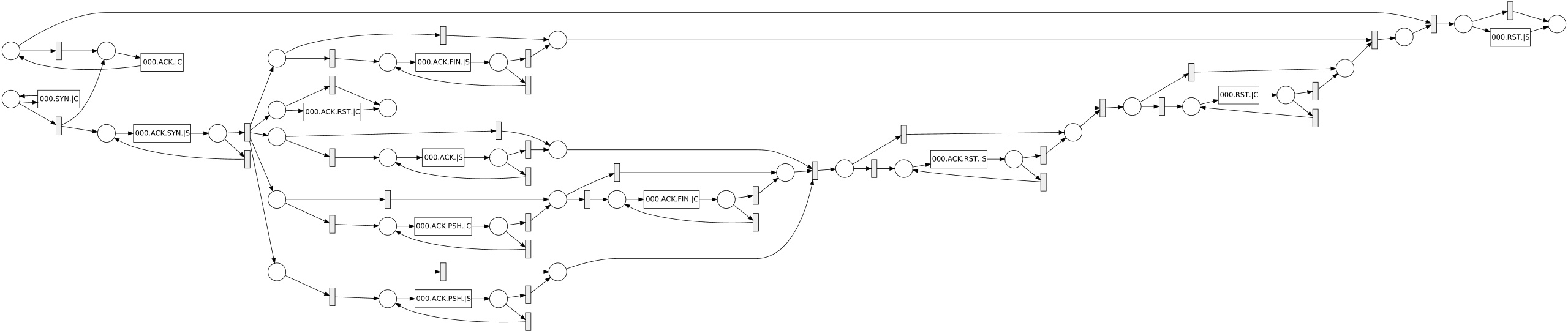}
        	\caption{The Petri nets mined with 100 cases.}
        	\label{fig:inductive_model_compare_100}
        	\end{center}
        \end{figure}

\begin{figure}[h]
        	\begin{center}
        	\includegraphics[width=4in]{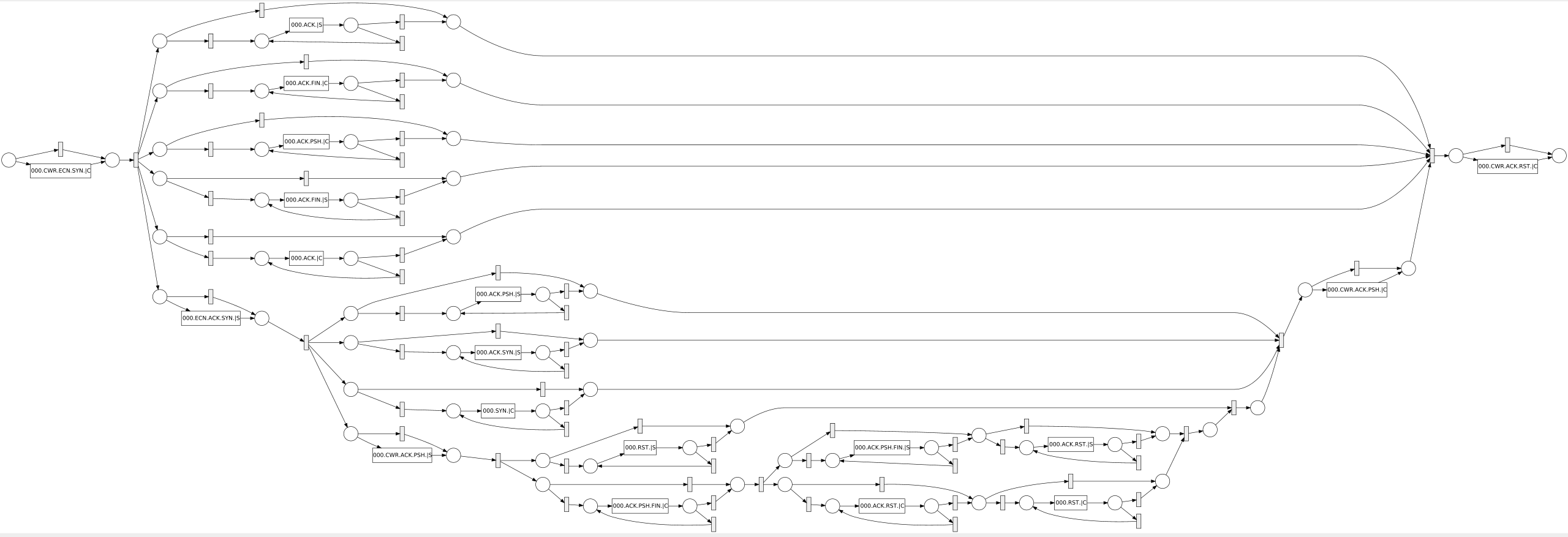}
        	\caption{The Petri nets mined with 20k cases.}
        	\label{fig:inductive_model_compare_20k}
        	\end{center}
        \end{figure}

        Now, let us compare process models mined using inductive miner for setups 1),  3) and 5) in Fig. \ref{fig:inductive_model_compare_5}, Fig. \ref{fig:inductive_model_compare_100} and Fig. \ref{fig:inductive_model_compare_20k} respectively. These setups represent the small, medium and large event logs. Event logs that are larger than 20k cases do not show significant difference in our experiment.
        
        The model mined with 5 cases is a small-log instance for testing how the number of cases impacts the generality and accuracy of the model. Nine different event classes are observed from these 5 cases, and the model mined from 20k cases contains 19 observed event classes. An \emph{event class} is essentially the name of an observed event; in our process model, it is the concatenated strings of both attributes from a case, for example, 000.SYN$|$C.
        
\begin{figure}[h]
        	\begin{center}
        	\includegraphics[width=4in]{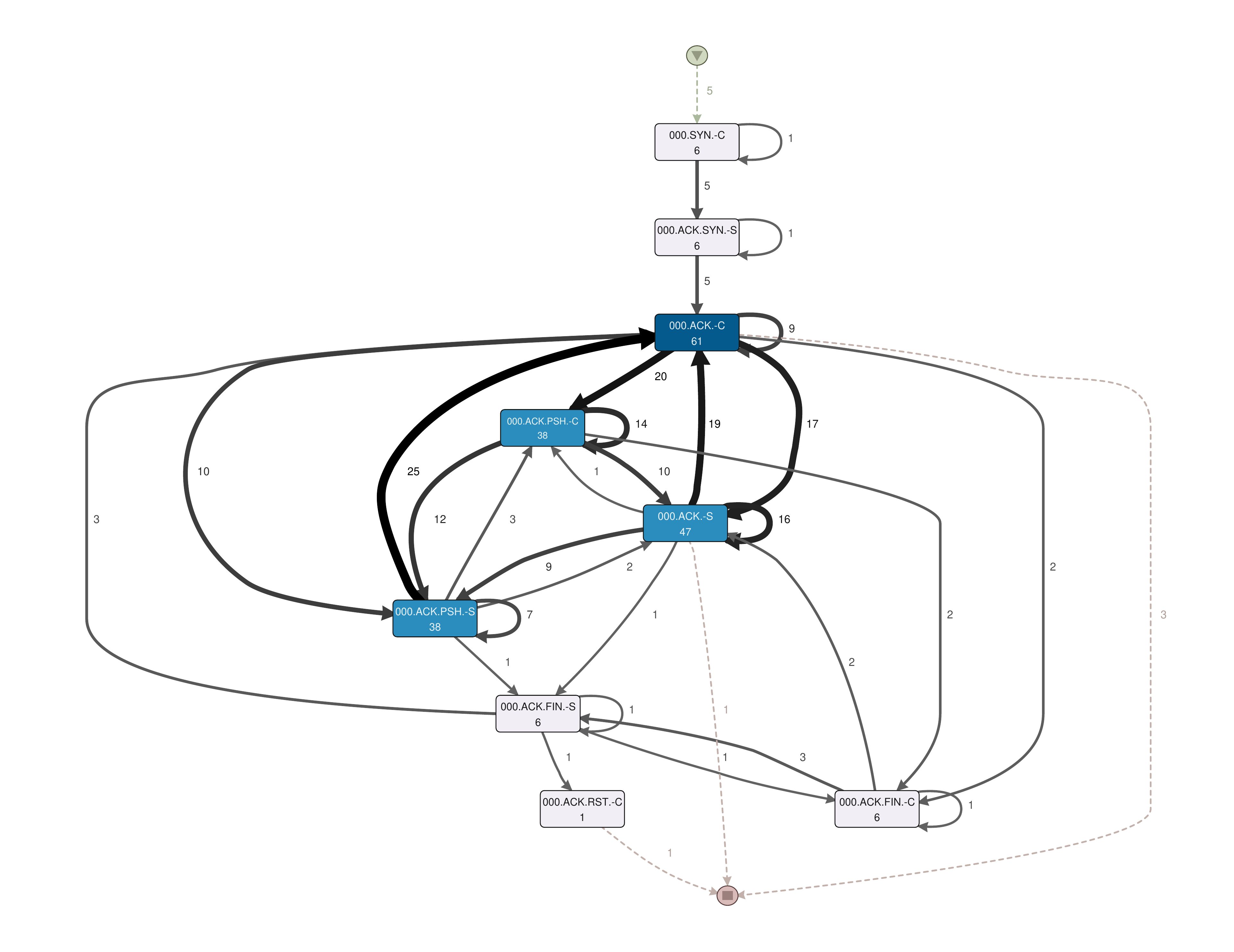}
        	\caption{The fuzzy models mined with 5 cases.}
        	\label{fig:fuzzy_model_compare_5}
        	\end{center}
        \end{figure}

\begin{figure}[h]
        	\begin{center}
        	\includegraphics[width=4in]{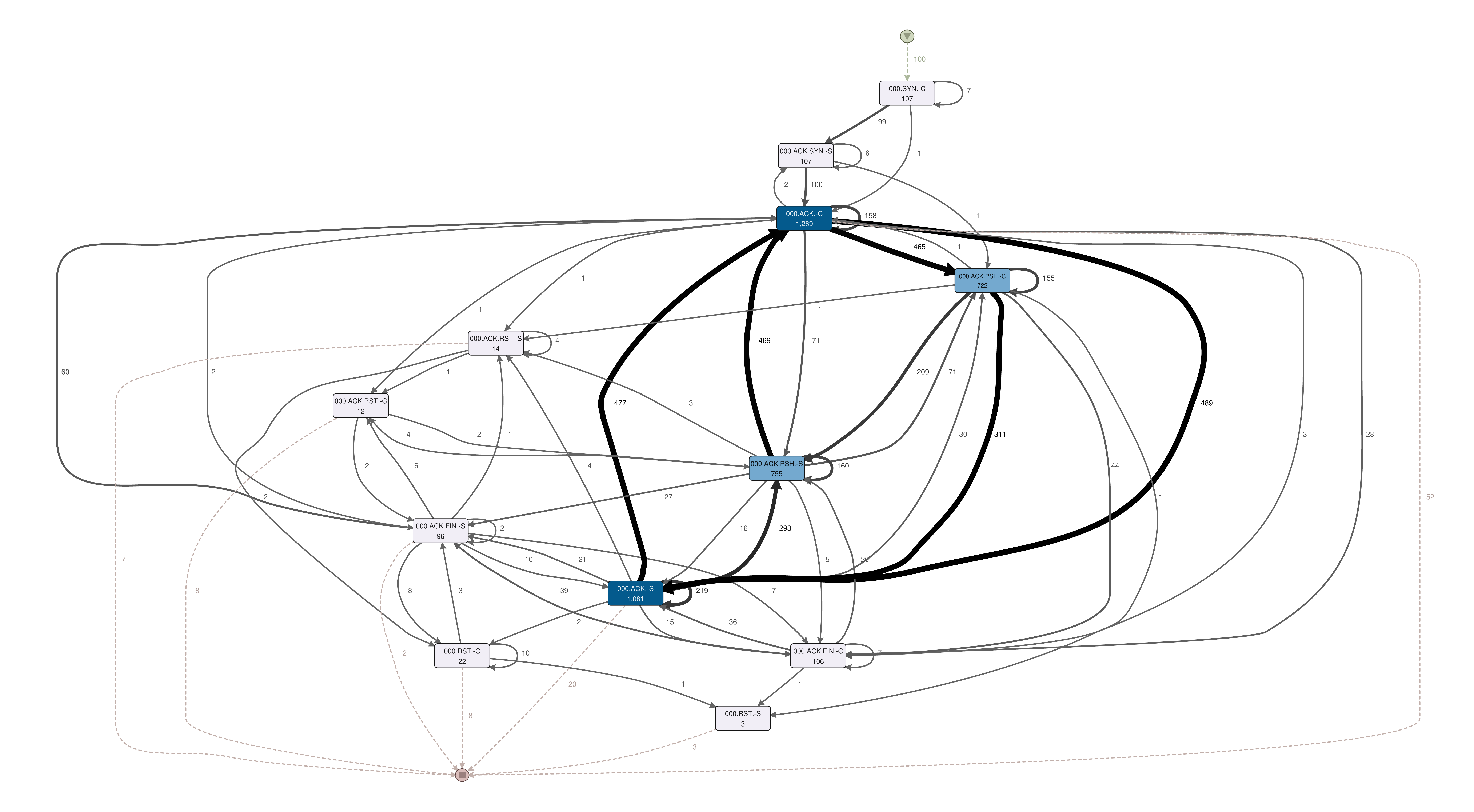}
        	\caption{The fuzzy models mined with 100 cases.}
        	\label{fig:fuzzy_model_compare_100}
        	\end{center}
        \end{figure}

\begin{figure}[h]
        	\begin{center}
        	\includegraphics[width=4in]{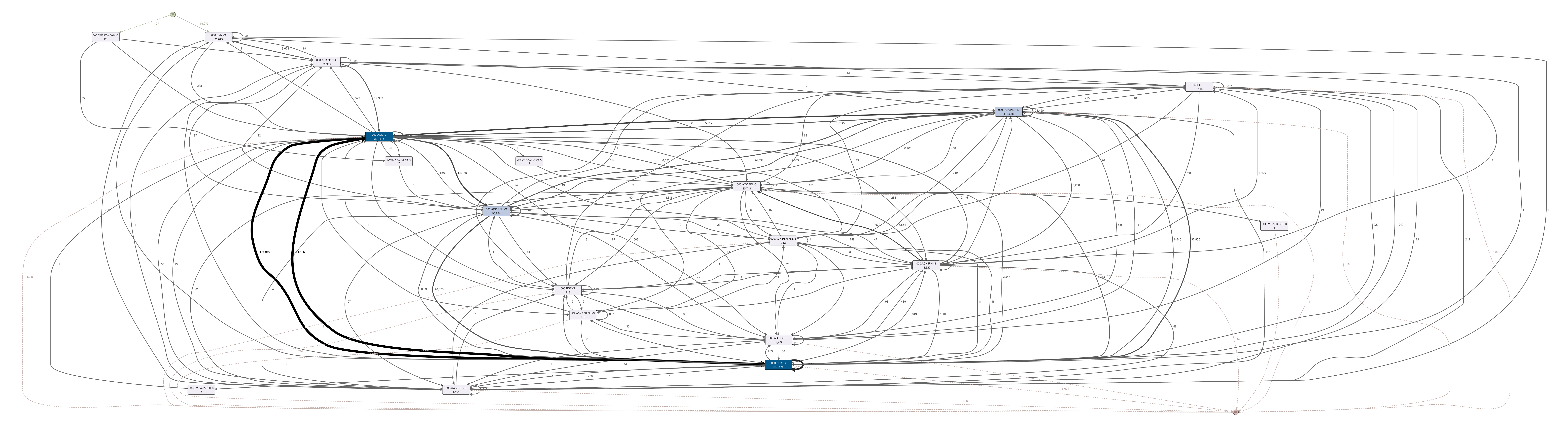}
        	\caption{The fuzzy models mined with 20k cases.}
        	\label{fig:fuzzy_model_compare_20k}
        	\end{center}
        \end{figure}

        From these two models, we see better accuracy on the model mined with 5 cases and better generality on the model mined with 20k cases. The model mined with 20k cases discovers the trace which terminates after the first packet of the three-way handshake (000.SYN$|$C); however, the same model also permits impossible transitions, such as entering the data-transmission stage without a three-way handshake. The model mined with 100 cases has a good trade-off between the models mined with 5 cases and 20k cases. We compare the process models mined with fuzzy miner with the same event logs from setup 1), 3) and 5) in Fig. \ref{fig:fuzzy_model_compare_5}, Fig. \ref{fig:fuzzy_model_compare_100} and Fig. \ref{fig:fuzzy_model_compare_20k} respectively.
        
        In our case, the fuzzy models provide a better balance between precision and generality for network traffic data. We evaluate the generality of small models and the precision of large models. The model mined with 5 cases is still quite accurate, similar to the inductive model; however, the differences between the models mined with 20,000 cases are greater. For instance, the model mined with 20,000 cases of Disco does not suffer from the handshake skipping issue. The fuzzy mining algorithm theoretically does not generate an edge unless the transition is observed in the event log. In other words, the large fuzzy model achieves greater generality, but retains a good accuracy.
        
    \subsection{Diagrams}
        Now we compare all process models with two state diagrams, the RFC TCP state transition diagram provided in \cite{postel1981transmission} and the improved diagram produces by Bishop et al. in \cite{bishop2005tcp}. We use this comparison for a better understanding of how TCP works, what PM can observe, and how the results of process mining are comparable with the common descriptions of TCP protocol by state diagrams. 
        
        \begin{figure}[h]
        	\centering
        	\includegraphics[width=3.1in]{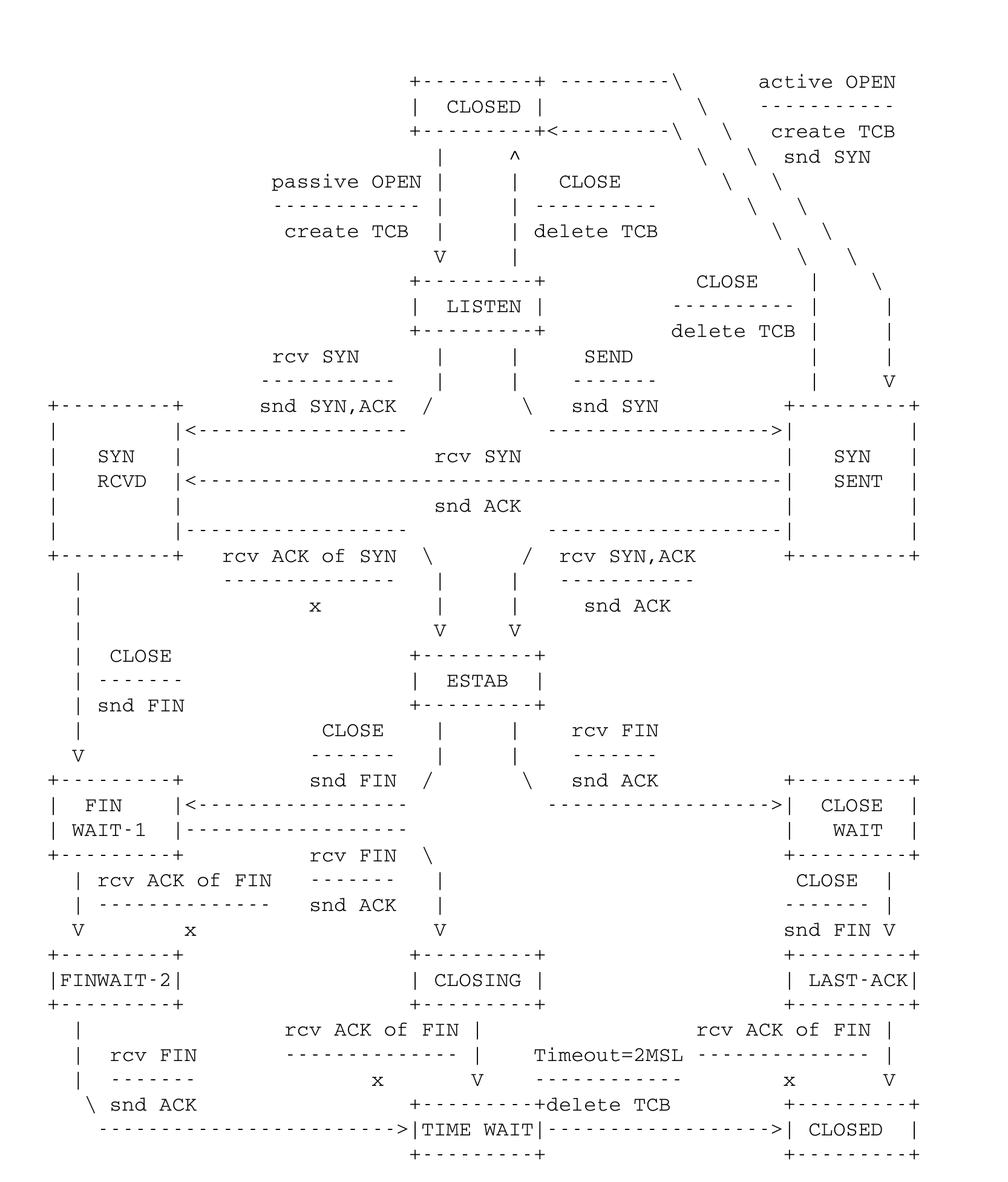}
        	\caption{RFC TCP state transition diagram \cite{postel1981transmission}.}
        	\label{fig:rfc}
        \end{figure}
        
        The RFC TCP state transition diagram describes a TCP flow in general terms. The diagram depicted the establishing, established, and closing stages of a TCP flow. This diagram depicts the establishing stage as a three-way handshake. The diagram lacks information regarding the established stage. The closing stage consists of two states on each side of communication (sending out FIN and waiting for ACK). This diagram has a much higher abstraction than any of the models mined, and all we can compare are the handshake and closing phases. In the diagram, cases such as disconnecting during a handshake are not permitted.
        
        \begin{figure}[h]
        	\centering
        	\includegraphics[width=3.1in]{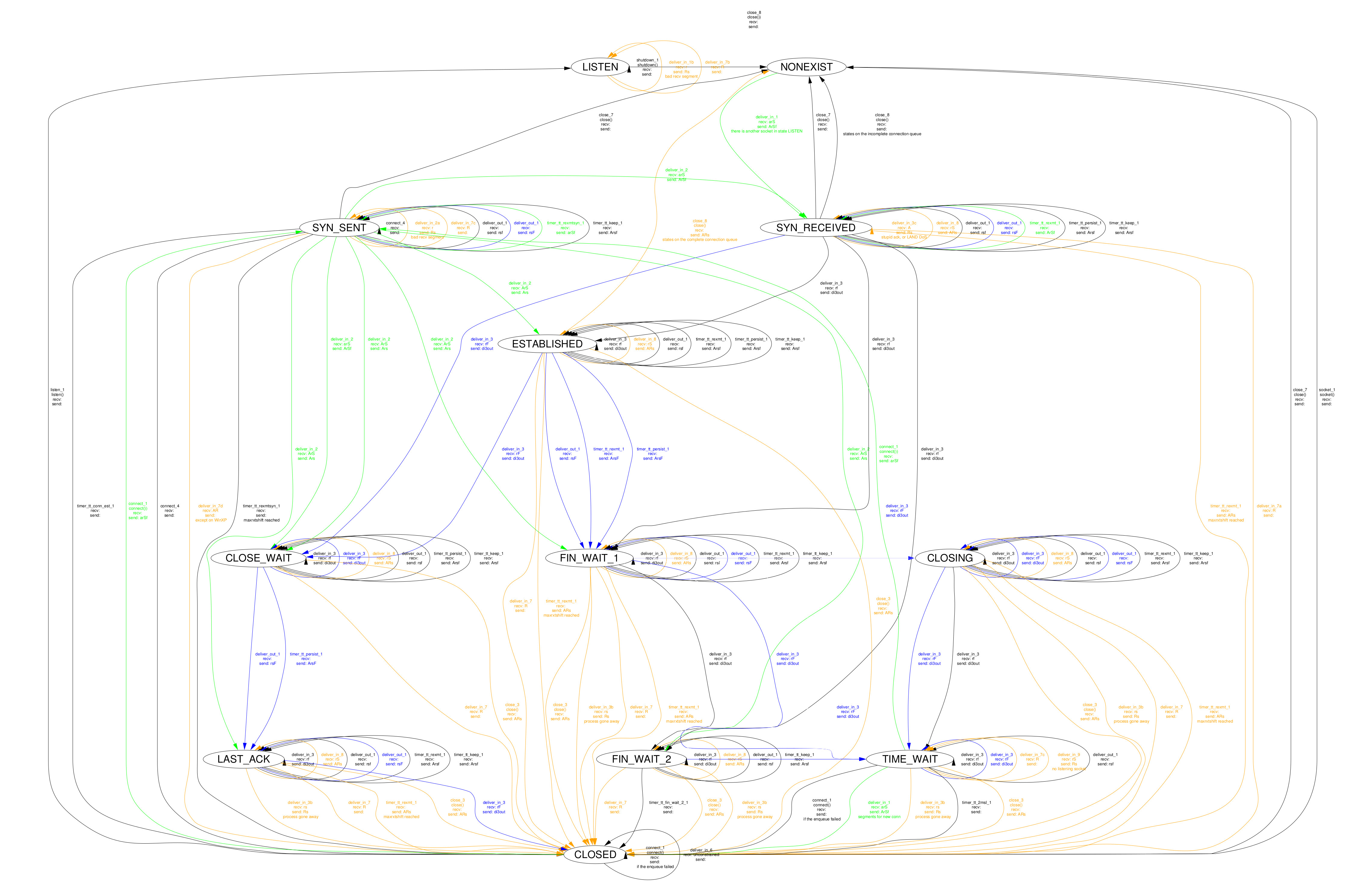}
        	\caption{State diagram by Bishop et al. \cite{bishop2005tcp}.}
        	\label{fig:bishop}
        \end{figure}
        
        The information shown by the state diagram of Bishop et al. is closer to the real traffic data as the analysis is based on real network traffic observed on the system calls. Transitions with RST set are coloured orange; transitions that have SYN set are coloured green; transitions that have FIN set are coloured blue; others are coloured black. We have flags in the diagram such as `Arsf', where `A' stands for ACK set, and `a' stands for ACK clear; `R' stands for RST set, and `r' stands for RST clear;`S' stands for SYN set, and `s' stands for SYN clear; `F' stands for FIN set, and `f' stands for FIN clear. We refer readers to the original report in \cite{bishop2005tcp} for more details.
        
        This state diagram has significantly more detail than the original RFC diagram. It displays additional traces and information about event classes containing the RST flag. START is not directly connected to PSH or END, similar to the process models discovered using Disco. Event names in our process models contain the labels S and C, which indicate whether a packet was sent upstream or downstream; however, the state diagram preserves the states of the RFC TCP state diagram, and it does not reflect where a packet was sent from.
        
        Based on our understanding of the the diagram, we make the transition matrix in Table \ref{tab:bishop_adj} that describe possible transitions and possible traces. For example, row 1 column 2 shows that after a packet with SYN set, we can possibly get a packet with ACK set next.
        
        \begin{table}[H]
            \caption{The adjacency matrix of state diagram by Bishop et al.}
            \centering
            \resizebox{0.8\textwidth}{!}{%
            \begin{tabular}{c|c|c|c|c|c|c|c|c|c|c|}
                     & SYN. & ACK. & ACK.SYN. & ACK.RST. & FIN. & DATA & ACK.FIN. & RST. & START & END \\ \hline
            SYN.     & 1    & 1    & 1        & 1        &      &      &          & 1    &       & 1   \\ \hline
            ACK.     &      &      &          &          &      & 1    &          & 1    &       &     \\ \hline
            ACK.SYN. & 1    & 1    & 1        & 1        & 1    &      &          &      &       &     \\ \hline
            ACK.RST. &      &      &          &          &      &      &          &      &       & 1   \\ \hline
            FIN.     & 1    & 1    &          &          & 1    &      & 1        & 1    &       &     \\ \hline
            DATA     & 1    & 1    &          & 1        & 1    &      & 1        & 1    &       &     \\ \hline
            ACK.FIN. & 1    & 1    &          &          & 1    &      &          &      &       &     \\ \hline
            RST.     &      &      &          &          &      &      &          &      & 1     & 1   \\ \hline
            START    & 1    &      &          &          &      &      &          & 1    & 1     &     \\ \hline
            END      &      &      &          &          &      &      &          &      &       &     \\ \hline
            \end{tabular}
            }
            \label{tab:bishop_adj}
        \end{table}
        
    \subsection{Overall Comparison}
        In Table \ref{tab:generality_comparison}, we compare the level of generality of various discussed models. A general model should permit normal, unobserved behaviours to pass compliance testing. Compared are only rare normal flows that exist in actual network traffic data. The examples illustrate some infrequent flows. The table columns display models, such as an inductive model mined with five traces, a fuzzy model mined with one hundred traces, etc. The row illustrates rare cases that we anticipate being observed in the models. If the case is observed in any process model or diagram, the cell is marked Y(es); otherwise, the cell is marked N(o). Following the inductive model in terms of generality is the fuzzy model. This result is anticipated due to the algorithmic design, in which an inductive miner explores patterns such as concurrencies, yielding a large number of possible traces that are not present in the event log. Alternatively, RFC diagrams are general descriptions of TCP states, and both diagrams are limited to predefined stages such as LISTEN, SYN-SENT, and ESTABLISHED, among others.
        
        \begin{table}[h]
        \caption{Generality Comparison.}
        \centering
        \resizebox{\textwidth}{!}{%
        \begin{tabular}{|c|c|c|c|c|c|c|l|c|c|}
        \hline
                                                & Ind. 5t & Fuzzy 5t & Ind. 100t & Fuzzy 100t & Ind. 20kt & Fuzzy 20k &  & RFC Diag. & Bishop \\ \cline{1-7} \cline{9-10} 
        PSH in established stage.       & Y       & Y        & Y         & Y          & Y         & Y         &  & N         & N      \\ \cline{1-7} \cline{9-10} 
        Duplicate pakages.                      & Y       & Y        & Y         & Y          & Y         & Y         &  & N         & Y      \\ \cline{1-7} \cline{9-10} 
        Reset during handskake.                 & N       & N        & Y         & Y          & Y         & Y         &  & N         & Y      \\ \cline{1-7} \cline{9-10} 
        Send of receive SYN after estanblished. & N       & N        & N         & N          & Y         & Y         &  & N         & N      \\ \cline{1-7} \cline{9-10} 
        Reset during closing                    & Y       & N        & Y         & N          & Y         & Y         &  & N         & Y      \\ \hline
        \end{tabular}
        }
        \label{tab:generality_comparison}
        \end{table}
        
        \begin{table}[h]
        \caption{Accuracy Comparison.}
        \centering
        \resizebox{\textwidth}{!}{%
        \begin{tabular}{|c|c|c|c|c|c|c|l|c|c|}
        \hline
                                                                                                                       & Ind. 5t & Fuzzy 5t & Ind. 100t & Fuzzy 100t & Ind. 20kt & Fuzzy 20k &  & RFC Diag. & Bishop \\ \cline{1-7} \cline{9-10} 
        Data transmission without handshake                                                                            & N       & N        & Y         & N          & Y         & N         &  & N         & N      \\ \cline{1-7} \cline{9-10} 
        RST or FIN before handshake                                                                                    & N       & N        & Y         & N          & Y         & N         &  & N         & N      \\ \cline{1-7} \cline{9-10} 
        Loop through Close and Established stages                                                                      & N       & N        & Y         & Y          & Y         & Y         &  & N         & N      \\ \cline{1-7} \cline{9-10} 
        \begin{tabular}[c]{@{}c@{}}Sending same packets infinitely without\\ response from the other side\end{tabular} & Y       & Y        & Y         & Y          & Y         & Y         &  & N         & N      \\ \hline
        \end{tabular}
        }
        \label{tab:accuracy_comparison}
        \end{table}

        \begin{figure}[H]
        	\begin{center}
        	\includegraphics[width=4.5in]{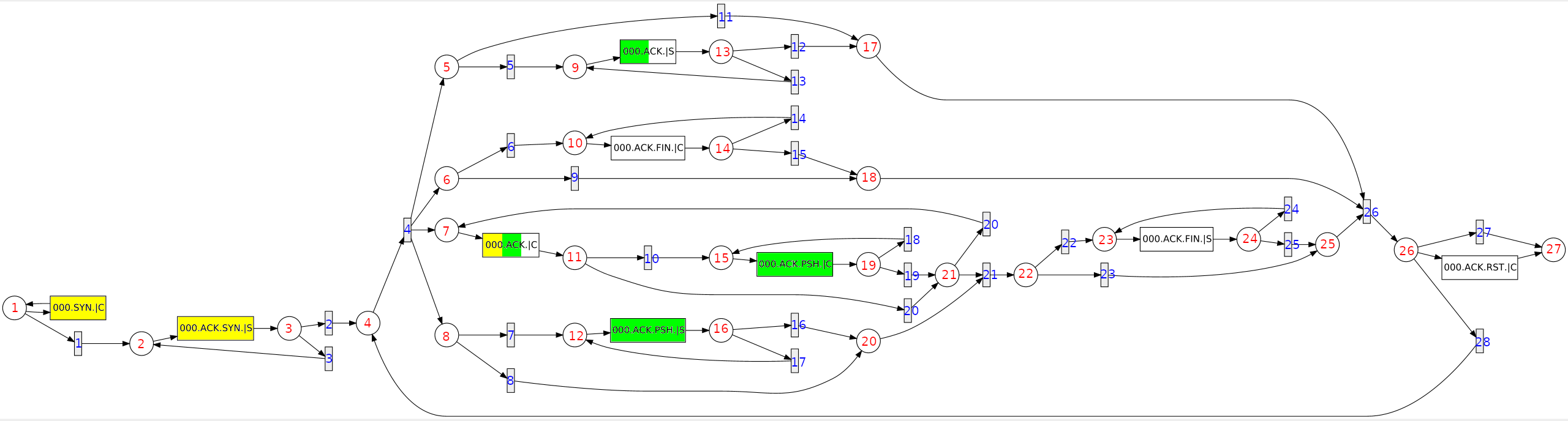}
        	\includegraphics[width=3.5in]{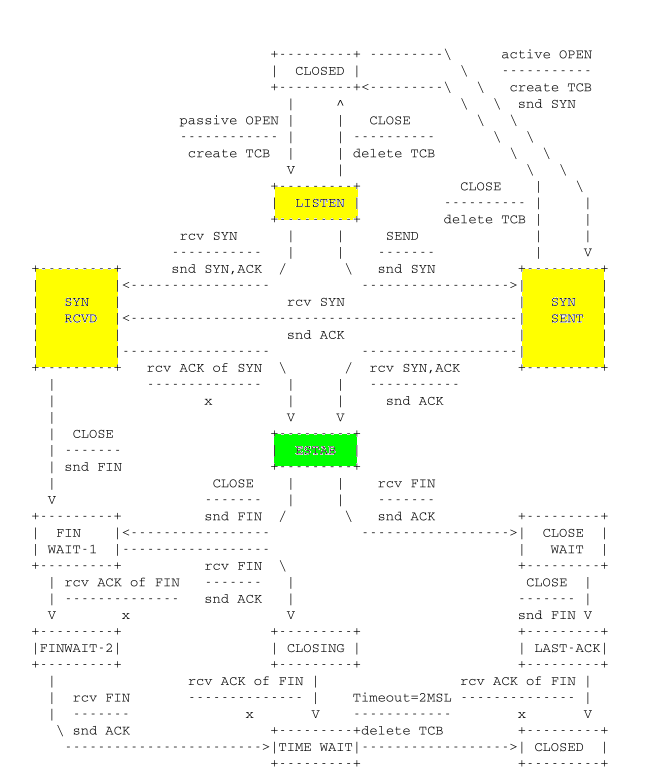}
        	\caption{Stage comparison between process model and diagram.}
        	\label{fig:stage_comparison}
        	\end{center}
        \end{figure}
        
        We contrast accuracy in Table \ref{tab:accuracy_comparison}. An accurate model that depicts normal behaviours should not allow abnormal traces to pass the compliance checking. We assume a trace is abnormal based on typical use cases because the behaviour of a flow can vary depending on the implementation of an application. In contrast to Table \ref{tab:generality_comparison}, we DO NOT anticipate any of the cases in Table \ref{tab:accuracy_comparison} to fit any process model or diagram. We achieve greater accuracy with fuzzy models, and the diagrams are the most accurate. Cases involving infinite loops are a common issue with process models. Because diagrams adhere to predefined stages, looping through the close stage and the established stage cannot occur. Due to the fact that ACK typically occurs after receiving any packet, a link will be created between this packet and the ACK packet, causing loops between stages. To better illustrate this limitation of process models, we colour the process model in Fig. \ref{fig:inductive_model_compare_5} into three stages; thus, we can directly compare it with the diagrams at the level of stages. Fig. \ref{fig:stage_comparison} depicts the coloured process model. The yellow nodes correspond to events from the three-way handshake stage; the green nodes correspond to events from the data transmission stage; and the uncoloured nodes correspond to events from the termination stage.
        
        The event class "000.ACK$|$C" occurs in all three stages of this example, whereas "000.ACK$|$S" occurs in the established and closing stages. This is the reason for the loops between stages being created, and the fuzzy model has the same problem. We label each packet with C or S to ensure that the information regarding which side of the terminals sent a particular packet will not be lost in PM, and by adding this information, we have also prevented the creation of additional loops. Although this information is not directly present in the packet data, it should be added to event logs by processing the raw packet data. Therefore, our experiment can also be improved by including information about the stages.
        
        On the basis of this observation, the question can be posed: what exactly can PM extract from event logs? In our case, without prior knowledge of the TCP protocol and preprocessing the packet data prior to adding them to the event log, the information of C/S and stages will be lost, resulting in a greater number of loops in mined modesl and a greater degree of difficulty for the analysis.
        
\section{Conformance Checking}
    \begin{table}[h]
    \caption{Fitness scores of conformance checking.}
    \centering
    \resizebox{\textwidth}{!}{%
    \begin{tabular}{|l|l|l|l|l|l|l|l|l|l|}
    \hline
    Normal & BruteForce & DoS & Heartbleed & CoolDisk & Dropbox & PortscanNmap & Web & Botnet\_ARES & Port\_Scan\_DDos \\ \hline
    1      & 1          & 1   & 1          & 1        & 1       & 1            & 1   & 1            & 1                \\ \hline
    \end{tabular}
    }
    \label{tab:conformance_checking}
    \end{table}
    
    Using the inductive models which is mined with a 20k event log (the same event log from setup \ref{model_for_comformance_checking}), we performed conformance checking and calculate the fitness score with ProM. A score of 1 indicates that all cases perfectly match the process model; otherwise, the score will be less than 1. The method for performing conformance testing is described in detail in \cite{rozinat2008conformance}. The average fitness scores for multiple setups are displayed in Table \ref{tab:conformance_checking}. Using a 20k-case event log, we mine the normal process model; then, we perform conformance checking with various categories including another normal log and all attacks. The normal column represents the fitness score of an additional subset distinct from the one used to generate the model. On Zenodo, all preprocessed data will be accessible\footnote{https://doi.org/10.5281/zenodo.6646875}.
    
    As all cases exhibit perfect model fit, it is impossible to identify the anomaly. Observable traces are a limitation of related works that mine the normal model from only a few dozens or hundreds of cases, resulting in a relatively accurate model. However, network flows can be significantly more complex; therefore, mining a process model from deviated flows will result in a significantly more general model. As long as the flows are being routed on the network, regardless of whether they are attacks or normal traffic, they always comply with TCP, and attacks typically do not exploit the protocol itself. We believe that the frequency distribution of transitions is more crucial for anomaly detection in IDS. An anomaly may not be viewed as a particular transitional change, but rather as the global frequency structure.
    
\section{Conclusion} \label{section:conclusion}
    We believe that using PM and conformance checking directly for network intrusion detection is ineffective. We compare models, and based on our observations, we know that it is difficult to detect anomalies due to the generality of the process model. In addition, we explained why using PM for network traffic anomaly detection is ineffective. A benefit of PM is that it encodes the global process structure, allowing parallelism to be discovered. This may be important for detecting distributed denial of service (DDoS), Botnet, and brute force attacks.

\let\url\nolinkurl
\bibliographystyle{splncs04}
\bibliography{main}

\end{document}